\documentclass{aa}
\usepackage{graphicx}
\begin{document}
\title{Unified model for quasar absorption line systems}
\author{Snigdha Das
\and Pushpa Khare
\and Archana Samantray}

\institute{Physics Department, Utkal University,
Bhubaneswar, 751004, India}

\date{Received date /Accepted date}

\abstract{
We propose a three component model consisting of minihalos and galactic
halos with embedded thin discs for absorbers producing all the observed
classes of intervening quasar absorption line systems. We show that
this model, based on CDM cosmology, can explain most of the observed
statistical distributions of various types of absorption systems.  Use
of the Schechter luminosity function for absorbers, on the other hand,
is consistent with the observations only if the number of galaxies was
larger in the past and reduced with time due to mergers. A strong
chemical evolution in the halos of galaxies is indicated by the
observed properties of CIV lines. We discuss our results in the light
of the recent observations of the absorption line systems.
}
\maketitle
\section{ Introduction}

Quasar absorption lines provide unique, sensitive and direct probes of
the gaseous content of galaxies and inter-galactic medium (IGM) over
look back times comparable to a Hubble time. Studies of these lines
promise to provide insight into the processes that regulate the
evolution of the IGM and the galaxies and the mechanisms that determine
the physical conditions in discs and in clouds embedded in the halo. In
a typical quasar spectra there are several populations of absorption
lines and it has become apparent that these different populations are
likely the absorption signatures of different regions and conditions
within the galaxies and the IGM.

Narrow, heavy element absorption systems (HEASs) in the quasar spectra
are believed to be produced by gas in the halos of intervening
galaxies. This was demonstrated by the high success rate of detection
of emission at the redshift of the intermediate redshift (z $\sim$
0.5-0.8) MgII absorption systems (Bergeron 1988; Steidel 1993).  High
resolution studies of these heavy element lines have revealed that the
lines break up into several distinct components which are most likely
produced by clouds in the halos of the absorbing galaxies (Srianand \&
Khare 1994). Most of the Lyman limit systems (LLSs), which have a
neutral hydrogen column density $>$ 2$\times 10^{17}$ cm$^{-2}$ are
associated with the HEASs.

Damped Ly$\alpha$ systems (DLASs) are produced by condensations of neutral
gas having neutral hydrogen column density N$_{\rm HI} > 10^{20.3} \rm
cm ^{-2}$ and are easily detectable in the spectra of background
quasars (Wolfe et al. 1986). These are associated with metal lines of
low ionization and often high ionized species as well. Although the
exact nature of the DLASs has yet to be understood it is
generally believed that they are produced by the progenitors of present
day galaxies. This conclusion is based on the following facts (i) the
mass contained in these systems at redshifts $>$2 seems to be
comparable to the stellar mass at present and (ii) the kinematics of
the gas in DLASs seems to indicate their origin in rotating discs. This
view has been questioned recently by several authors in the light of
some of the new data on these systems. We will discuss these later on
in this paper.

Lanzetta et al. (1995) and Chen et al. (1998), from a comparison of
galaxy and absorber redshifts showed that a dominant fraction of low
redshift (z$<$1) Lyman alpha absorption systems (LAASs) arise in
extended halos of galaxies. This was also corroborated by the discovery
of heavy elements (Cowie et al. 1995) at redshifts around 3 and of
clustering in these systems at redshifts between 1.7 and 3.8 (Srianand
\& Khare 1994). Recently Chen et al. (2000) have considered the
statistical properties of the high redshift galaxies in the Hubble deep
field and the statistical properties of Ly $\alpha$ absorption
systems.  They conclude that at least 50\% and possibly as high as
100\% of observed LAASs arise in extended envelopes of galaxies at
z$>$2.  It thus appears that a substantial fraction of the LAASs may
be produced by the outer parts of galactic halos rather than in
the IGM as believed earlier (Sargent et al. 1980).

During the past few years, there has been a maturing of attempts to
understand the nature of these intervening population, evolving from
rather simple idealized models (e.g. pressure confined clouds,
gravitationally confined "minihalos", freely expanding clouds etc) to
sophisticated numerical simulations that are able to reproduce many of
the features of the absorption spectra in the context of the evolution
of structure (Chiba \& Nath 1997; Das \& Khare 1999; Charlton \&
Churchill 1996; Kauffmann 1996; hereafter K96, Gnedin \& Ostriker, 1996
etc.). Das \& Khare (1999) have considered a two component (minihalos
and galactic halos) CDM model to understand the LAASs and associated
CIV absorption lines.  K96 considered the formation of disc in CDM
models and showed that the observed properties of the DLASs can be
accounted for by these models.  This pure disc model for DLASs was
questioned by Prochaska \& Wolfe (1997) on the basis of kinematic
structure of the DLASs.  Subsequently Jedamzik \& Prochaska (1998) have
shown that only a finely tuned disc model within the framework of CDM
could be made marginally consistent with the damped Ly $\alpha$
observations.

The purpose of this paper is to explore a consistent model of minihalos
and of galaxies which are assumed to have spherical and photo ionized
gaseous clouds embedded in their halos with a disc at the centre,
to better understand the nature of the LAASs, HEASs  and DLASs in a
unified manner. In section 2, we describe the basic model with the
underlying hypothesis. We discuss the orientation of the discs inside
the halo in section 3. We explore the field of parameters and constrain
these parameter using observations in section 4. We present the results
in section 5 and 6 before drawing our conclusions in section 7. A
discussion of the results in view of some of the recent observations is
presented in section 8.

\section{ The basic model}

The two component model was recently considered by Das \& Khare (1999)
to understand the observations of Lyman $\alpha$ forest lines and their
accompanying C IV lines in the framework of the hierarchical structure
formation model in CDM cosmology. This model consists of minihalos and
galactic halos. These are dark matter halos having 5\% of mass in the
form of baryons. The Minihalos have circular velocities, V$_{\rm c}$
between 15 and  55 km s$^{-1}$ and consequently have smaller virial
radius (see eqn 2.1). The lower limit of 15 km s$^{-1}$ is roughly the
value below which the perturbations in the baryonic mass are suppressed
while 55 km s$^{-1}$ is roughly the value below which the
photo ionization by the external UV background suppresses cooling and
therefore the formation of clouds and stars. These are chemically
enriched if at all, by an earlier generation of stars. 

Halos with circular velocities larger than 55 km s$^{-1}$ are called
galactic halos. Upper limit on V$_{\rm c}$ for galactic halos is taken
to be 500 km s$^{-1}$ as the results presented below do not change if
the upper limit is increased above this value. The gas in these halos
can cool due to higher gas densities in here (see below) and form
clouds and stars.  Thus these halos have clouds embedded in it which
are in pressure equilibrium with the hot intercloud medium. Their
number density is assumed to have the same radial dependence as the
total density.  Here, we extend this model to include the presence of
isothermal discs inside the galactic halos which form as the gas cools,
collapses and forms stars (K96). It may be noted that all the halos in
our model, being gravitationally confined are assumed to be spherical.
There have been suggestions in the literature of the filamentary nature
of the Ly $\alpha$ absorbers (Charlton et al 1993).  These were based
upon the large transverse sizes of these absorbers ($\ge$ 100 kpc)
obtained from the occurrence of common absorption lines in the spectra
of multiple images of the lensed QSOs (Smette 1995). Similar sizes
along the line of sight (los) (as would be required for the spherical
absorbers) would require too low a value of the external (IGM)
pressure. This is however, not a problem in our model as the galactic
halos (where the lines observed by Smette (1995) at low resolution
would have originated) have clouds with sizes of about a few kpc
embedded in them. The common absorption lines in the images of lensed
QSOs may be produced by different clouds in the same halo.  Thus the
assumption of spherical halos does not violate the constraints from the
lensed QSO data. We note that the lensing data indicate that if the 
los to one image intercepts an absorber the probability that the los to 
the other image intercepting the same absorber is $>90 \%$. We will 
discuss this with reference to our model in section IV.

Our model for the disc is based on the work of K96 according to which
the gas collapses to a constant fraction of its initial virial radius
to form a rotationally supported disc within the halos, the virial
radius for spherical collapse model is given by

\begin{equation}
\rm r_{\rm vir} = 0.1\rm H_{\rm o}^{-1}(1+\rm z)^{-3/2} \rm V_{\rm c}.
\end{equation}
Here H$_{\rm o}$ is the Hubble constant. Using the results of N body
simulation (Barnes \& Efstathiou 1987) it can be shown that for a
disc/halo mass ratio of 0.1 the equilibrium disc radius is $\simeq$
10\% of the virial radius of the halo. A spread around this factor is
expected due to possible spread in spin parameters of the dark halo
(K96). We have, included an additional dependence of the disc size on
redshift apart from the (1+z)$^{-3/2}$ dependence which is due to the z
dependence of r$_{\rm vir}$. This seems to be necessary to account for
the observations as seen below. Thus the disc radius in our model is
given by,

\begin{equation} \rm R_{\rm d} = 0.1\rm H_{\rm o}^{-1}(1+\rm z)^{\rm a
-3/2} \rm V_{\rm c},\end{equation} where a is an unknown parameter
whose value as well as the value of the thickness of the disc, t, is fixed
by comparing the model results with observation of DLAs as explained
below.

Following K96, we assume all the discs to have flat rotation curves and
a rotational velocity equal to the circular velocity of their
surrounding dark matter halos. Assuming Kennicutt (1989) star formation
law and a constant gas velocity dispersion of 6 km/s the gravitation
stability condition gives the critical surface density as (K96)

\begin{equation}\Sigma_{\rm crit}(\rm R)=0.59 {{\rm V(\rm km \;\rm
s^{-1})}\over {\rm R(\rm kpc)}}  M_\odot\rm pc^{-2},\end{equation}
where V is the velocity of the disc and R is the radial distance from
the centre. The surface gas density in the disc remains close to the
critical density with constant in fall, star formation and ejection of
hot disc gas into the halo through supernovae explosions. So the
surface gas density in the disc is also given by the above equation.

The total hydrogen column density contributed by the disc for an impact
parameter P for a face on disc can be written as,

\begin{equation} \rm N_{\rm HT}^{\rm disc} (\rm P) =  {{\Sigma (\rm
P)}\over {\rm m}}, \end{equation} m being the average mass per gas
particle which we assume to be mass of the proton, m$_{\rm p}$. The
average number density, n$_{\rm H}$ along a los through
the disc is thus

\begin{equation} \rm n_{\rm H} = {{\rm N_{\rm HT}^{\rm disc} (\rm
P)}\over {\rm t}}.\end{equation} The total column density along a given
los is the sum of the column density contributed by the disc, the
column densities in all the clouds in the halo that lie along the los
and the column density contributed by the hot medium along the los. We
assume the number density of clouds at a radial distance r from the
centre of the halo, $\eta_{\rm cl}$(r) to be

\begin{equation}\rm
\eta_{cl}(r)\;={\eta_{cl}(0)\over{1+r^{2}/r^{2}_{c}}},\end{equation}
r$_{\rm c}$ being the core radius. The number of clouds per unit volume
at the centre of the galaxy is given by \begin{equation}\rm {\eta_{\rm
cl}(0) }= \rm f\;{{3}\over{4\;\pi\;r^{3}_{cl}(0)}}.\end{equation} The
value of f is fixed by comparing the model results with observations,
as described below.

The radial density distribution of the baryonic matter in the minihalos
and in galactic halos is given by

\begin{equation}\rho (\rm r)\;={{\rm f_{b}\rm V_c^{2}}\over{4\;\pi\;\rm
G\;(r^{2}+r_c^{2})}}\;({\rm x_v\over{\rm x_v-\rm arc \;\rm tan\; \rm
x_v}}),\end{equation}
where $\rm f_{b}$ is the fraction of baryonic matter in the halo, 
which we take to be 0.05 as mentioned above.
r$_{\rm c}$ is the core radius which is taken to be 10 and 100 kpc for
mini and galactic halos respectively and $ \rm x_{\rm v} ={\rm r_{\rm
vir}\over{\rm r_{\rm c}}}\simeq 15$ (Chiba \& Nath  1997).

\section {\bf {Orientation of the discs}} 

The orientation of the disc is an important factor while determining
the column density contribution of the disc which is likely to increase
due to the increased path length through an inclined disc.  For the
calculation of the contribution of the disc to the column density along
a los with impact parameter P, we assume the disc to be randomly
oriented with inclination i. At this orientation, the disc is projected
on the plane of the sky on to an ellipse with area $\pi{\rm R_{\rm
d}}^{2}\rm cos\;\rm i $. If the los is located at a random angle
$\theta$ from the major axis of the ellipse then the radial position
along the surface of the disc where it enters the plane of the disc is
given by

\begin{equation} \rm R_{\rm in} =  \rm P [ \rm cos ^{2} \theta  +
({{\rm sin ^{2} \theta}\over {\rm cos \;\rm
i}})^{2}]^{1/2},\end{equation} and the radial position along the
surface of the disc where it leaves the plane of the disc is given by

\centerline{$ \rm R_{\rm out}$ = [ P$^{2} \rm cos^{2} \theta$ + ( P
${\rm sin\; \theta}\over {\rm cos\;i}$ + t tan i)$^{2}$]$^{1/2}.$} The
column density for a particular ion along a los with impact parameter
P, for minihalos is given by,

\begin{equation}{\rm N_{ion}(P)}\;=2{\rm{\int\limits_{P}^{\rm R_{\rm H}}{{\rm n_{\rm
H}(\rm r) \rm f_{\rm ion}(r) r dr}\over{\sqrt{\rm r^{2}-\rm P^{2}}}}}}.\end{equation}
Here $\rm n_{\rm H}(\rm r)$ is the number density of hydrogen, $R_{\rm H}$ is the radius of the halo 
and $\rm f_{\rm ion}(\rm r)$ is the ionized fraction of the ion
multiplied by Z, the abundance of the element being considered.
f$_{\rm ion}$ depends on the ionization parameter $\Gamma$ at the point
under consideration. For galactic halos, when the los passes outside
the disc the column density is given by

\begin{equation}{\rm N_{ion}(P)}\;=2{\rm{\int\limits_{P}^{R_{H}}{\rm
A(\rm r)dr}}}\; +\; 2{\rm{\int\limits_{P}^{R_{H}}{\rm B(\rm r)
dr}}}.\end{equation}
Here, A(r) = ${N_{\rm H}(\rm r) \rm f_{\rm ion}(r)
\sigma(r)\eta_{cl}(r)r}\over{\sqrt{\rm r^{2}-\rm P^{2}}}\;$ determines
the contribution from the galactic clouds, $\sigma(\rm r)$ being the
cross section of the clouds. $N_{\rm H}(\rm r)$ is the total hydrogen
column density contributed by an individual cloud present at the radial
distance r from the halo centre. B(r) = ${{\rm n_{\rm H}(\rm r) \rm
f_{\rm ion}(r) r }\over{\sqrt{\rm r^{2}-\rm P^{2}}}}$ is the
contribution to the column density from the intercloud medium.

When the los with an impact parameter P passes through the disc, the
column density for a particular ion is,

$${\rm N_{ion}(P, \theta, \rm i)}\; =
{\rm{\int\limits_{R_{H}}^{\sqrt{R_{\rm in}^{2} + (t/2)^{2}}}}}{\rm
A(\rm r) dr}\; +\; {\rm{\int\limits_{\sqrt{R_{\rm out}^{2}\; +
(t/2)^{2}}}^{R_{H}}}}{\rm A(\rm r) dr}\; +\;$$\begin{equation}
{\rm{\int\limits_{R_{H}}^{\sqrt{R_{\rm in}^{2}\; + (t/2)^{2}}}}}{\rm
B(\rm r) dr}\; +\; {\rm{\int\limits_{\sqrt{R_{\rm out}^{2}\; +
(t/2)^{2}}}^{R_{H}}}}{\rm B(\rm r) dr}\; +\;{{\rm f_{\rm ion} \Sigma(P)}
\over {\mu \rm m_{\rm p}}}.\end{equation}
The first two terms give the contribution of the clouds in the halo
along the los, the next two terms give the contribution from the
intercloud medium in the halo while the last term gives the
contribution from the disc. The disc contribution to the column density
along the sides of the projected ellipse is constant, but the halo
contribution changes from point to point. Here $\mu = \rm cos \rm i$
and the factor of ${1} \over {\mu}$ accounts for the increased path
length through the disc due to its inclination;  ${\rm f_{\rm ion}
\Sigma(P)} \over {\rm m_{\rm p}}$ is the column density contributed by
the disc in the face on position.

The ionization parameter $ \Gamma$ at a radial distance r from the
centre is given by $\Gamma(\rm
r)\;={{4\;\pi\;10^{-21}J_{21}\over{c\;\alpha_{\rm Q}\;\rm h\;n_{\rm H}(\rm
r)}}}$ where $\alpha_{\rm Q}$ is the slope of the spectrum of the UV
background radiation (assumed to be a power law). We assume
$\alpha_{\rm Q}$ = -1.5.  J$_{-21}$ is the intensity of the UV background
radiation in units of 10$^{-21}$ erg cm$^{-2}$ s$^{-1}$ Hz$^{-1}$
str$^{-1}$ at 1 Rydberg. c and h are the velocity of light and the
Planck's constant respectively.  The photo ionization code (cloudy 90)
was run for a range of hydrogen column densities and a table  ${\rm
f_{\rm ion}\over{Z}}$ vs. $\Gamma$ was prepared for each value of
hydrogen column density. These values are quite insensitive to the
values of particle densities and abundances. This table was used to
determine the value of ${\rm f_{\rm ion}\over{Z}}$ for any value of
$\Gamma$ and N$_{\rm H}$ during the calculations.

The column density distribution function for a particular ion produced
by halos with circular velocity $\rm V_{\rm l} < \rm V_{\rm C} < \rm
V_{\rm u}$ is given as,

$${\rm d^{2}N(z,\rm N_{ion})\over{d\rm N_{ion}dz}}\;= {{\rm c\over{\rm
H_{o}}}\;(1+\rm z)^{1/2} {\rm{\int\limits_{\rm V_{l}}^{\rm V_{u}}\rm n(\rm
V_{c},z)\;\epsilon\;\rm dV_{c}}}}$$
$$\hspace{1.5 in} {\rm{\int\limits_{\rm 0}^{\rm 1} d\mu |\frac{\partial \rm A(N_{ion},\mu)}{\partial \rm
N_{ion}}|}} $$

$$\;\;\;\;\;\;\;\;= {\rm c\over{\rm H_{o}}}\;(1+\rm
z)^{1/2} {\rm{\int\limits_{\rm V_{l}}^{\rm V_{u}}\rm n(\rm
V_{c},z)\;\epsilon\;\rm d V_{c}\;}}$$
\begin{equation}\hspace{1.3 in} {\rm{\int\limits_{0}^{1}\rm d
\mu\;\frac { \partial  }{\partial \rm N }\;}} {\rm{\int\limits_{\theta
= 0}^{2\; \pi}\rm d \theta\;{1 \over 2 }\rm
P^2(\theta,\mu)\;\;}},\end{equation} where $|\frac{\partial \rm
A(N_{ion},\mu)}{\partial \rm N_{ion}}| = {{{\rm A(\rm N_{ion} + \Delta
\rm N_{\rm ion},\mu)} - {\rm A(\rm N_{ion},\mu)}} \over {\Delta \rm
N_{ion}}} $ and area $ \rm A(\rm N_{\rm ion},\mu) =
{\rm{\int\limits_{\theta = 0}^{2\; \pi}\rm d \theta\;{1 \over 2} \rm
P^2(\theta,\mu)\;\;}}   $ is the projected area, for a given value of
i, inside which the column density is $> \rm N_{ion}$ and n(\rm V
$_{\rm c}$,\rm z)dV$_{\rm c}$ is the number of halos per unit volume
with circular velocity between V$_{\rm c}$ and V$_{\rm c}$ + dV$_{\rm
c}$ at redshift z. The deceleration parameter q$_{\rm o}$ is taken to
be 0.5 . We have chosen flat models for the universe ( q$_{\rm o}$=0.5)
which are popular (Ostriker \& Steinhardt, 1995) as they can be
produced naturally in any inflationary scenario and require no fine
tuning of parameters.  These models are also consistent with
independent constraints from various cosmological observations.
$\epsilon$ is the fraction of halos that give rise to absorption lines;
for the galactic halos it is the fraction of halos having sufficient
gas to produce absorption lines.  For these, $\epsilon$ is taken to be
0.69, which is the fraction of late type galaxies (Postman \& Geller
1984).  For mini-halos $\epsilon = 1$.  The redshift distribution of
lines of a particular ion with column density greater than N$_{\rm
ion}$ is given by

$${\rm dN(z,\rm N_{ion})\over{dz}}\;= {{\rm c\over{\rm
H_{o}}}\;(1+\rm z)^{1/2} {\rm{\int\limits_{\rm V_{l}}^{\rm V_{u}}\rm
n(\rm V_{c},z)\;\times}}}$$
\begin{equation}\hspace{1.3 in}{\rm{\int\limits_{0}^{1}
\rm d \mu\  \rm A(\rm N_{ion},\mu)\;\epsilon\;\rm dV_{c}\;}}. 
\end{equation} We
assume two models for the velocity distribution of halos : \newline (i)
Mass condensations in the CDM universe obtained using the
Press-Schechter theory (hereafter PS distribution) :  From the CDM
models of structure formation (Mo, Miralda-Escude \& Rees 1993) the
mass and circular velocity of a halo are related to the comoving radius
r$_{\rm o}$ and redshift z by,
 
\begin{equation}\rm M\;=\;{4\;\pi\over{3}}\;\rho_{o}\;\rm
r_{o}^{3},\;\;\;\;\rm V_{c}\;=\;1.67\;(1+\rm
z)^{1/2}\;H_{o}\;r_{o},\end{equation} where $\rho_{\rm o}$ is the mean
density of the universe at the present time.  The number of halos per
unit volume with circular velocity between V$_{\rm c}$ and V$_{\rm c}$
+ dV$_{\rm c}$ is given by 
$$ \rm n(\rm V_{\rm c},\rm
z)\rm dV_{\rm c}\;=\;{-3(1.67)^{3}\delta_{c}\rm
H_{o}^{3}(1+z)^{5/2}\over{(2\pi)^{3/2}V_{c}^{4}\Delta (r_{o})}}\;{\rm
d\;\rm ln\;\Delta\over{\rm d\;\rm ln\;V_{c}}}\;$$
\begin{equation}\hspace{1.3 in} \times\;\rm
exp({-\delta_{c}^{2}(1+z)^{2}\over{2\;\Delta^{2}
(r_{o})}})\;dV_{c}.\end{equation} Here $\delta_{\rm c}$ = 1.68 and the
functional form of $\Delta({\rm r_{o}})$ for the CDM power spectrum of
density perturbation is \begin{equation} \Delta(\rm
r_{o})\;=\;16.3\;\rm b_{\rm g}^{-1}\;(1-0.3909\rm
\;r_{o}^{0.1}+0.4814\rm \;r_{o}^{0.2})^{-10},\end{equation}
where b$_{\rm g}$ is the bias parameter which is taken to be 1 in our
calculations.\newline (ii) Observed (Schechter) luminosity function of
galaxies (hereafter LF distribution) :  The number of spiral galaxies
per unit volume, with luminosity between L and L+dL is given by,

\begin{equation} \rm n(L)dL = \phi ^{*} ({\rm L\over{\rm L_{*}}})
^{\alpha } exp(-{\rm
L\over{L_{*}}})\rm dL\;.\end{equation} 
Here $\phi ^{*}$ and $ \alpha$ give the normalization and the index of
faint end slope of the luminosity distribution. L$^{*}$ is the
luminosity of the galaxies having circular velocity V$_{\rm c}^{*}$ which
we take as 250 km s$^{-1}$. The luminosity L is related to circular
velocity V$_{\rm c}$ as ${\rm L\over{\rm L_{*}}} = ({\rm V_{\rm
c}\over{\rm V_{\rm c}^{*}}})^{2.6}$ for spiral galaxies. Bromley et
al. (1998) have determined the values of the luminosity function
parameters for six spectral type of galaxies in two different
environments. These values are used here  to calculate the total number
of galaxies as a function of the circular velocity and redshift.

It is possible that the number density of galaxies has been higher in
the past and reduced with time due to merger of galaxies. We also
consider this possibility. The number density of galaxies after merging
is given by n($\delta \rm t$) = f$(\delta \rm t) \rm n_{\rm o}$ (Park
\& Gott 1997) where n$_{\rm o}$ is the present number density of
galaxies and $\delta \rm t $ is the look back time of the universe
given by, $\delta \rm t = {2\over 3}(1-(1+z)^{-1.5})$. We assume that
there is no evolution of the velocity dispersion. The strength and time
dependence of merging is described by the function f($\delta \rm t$)
which is assumed to be exp$(\rm Q (\delta \rm t))$ .  The merging rate
Q roughly corresponds to the average number of pieces at z $ \sim $0.65
that merged to produce one galaxy at z = 0. The value of Q varies
between 2 to 4 for different models of the universe; here we take Q=3.5
as that gives the best match between model predictions and
observations.

\section {\bf {Constraining the model parameters}} 

The free parameters in our model are $\eta_{\rm cl}(0)$, the number
of clouds per unit volume at the centre, thickness of the disc, t and
a, which determines the rate of evolution of the disc size.
Increasing t decreases the particle density and therefore decreases the
fraction of neutral hydrogen. A given value of N$_{\rm HI}$ is thus
produced at smaller values of the impact parameters for larger values
of t reducing the number of DLAs per unit redshift. Change in a on the
other hand changes the slope of the redshift distribution of the DLAs.

The observed number density of DLA absorbers as a function of redshift
has been obtained by Lanzetta et al. (1995) and Storrie-Lombardi et
al. (1996).  The number density evolution of DLASs with N$_{\rm HI} >
20.3$ for our model is shown in Fig.1 for 4 values of a and three
values of t for PS distribution and a=0.9 and t=0.01 R$_{\rm d}$ for LF distribution. We find
that the model results for the redshift distribution of DLAs match best
with the observations for redshifts smaller than 4 for t $\simeq$ 0.01 R$_{\rm d}$
and a $\simeq$ 0.9.  These are the values we use in the rest of the
calculations.  We however, note that the observed number of DLASs at z
$\sim$ 4 is much larger than the prediction of all our models. Note
that as the DLASs are formed purely by the disc, their number does not
depend on $\eta_{\rm cl}(0)$.

Storrie - Lombardi et al. (1994) have studied that the redshift
distribution of LLSs and have found it to be roughly a power law with
slope 1.55 over the redshift range 0.40 to 4.69. However, they find
that the distribution becomes steeper for redshifts z $>$ 2.5. Their
values are plotted in Fig.2. The figure also shows the results of our
calculations for the number density evolution of LLSs for two values of
f (which determines the number of clouds per unit volume at the centre
of the galaxy, see eqn 2.7) and for the halo radius R$_{\rm H}$ = 1000
and 500 kpc for PS distribution. The radius of the halo does not make
any difference to the distribution.  This is because the contribution
from the outer halo to HI column density is negligible for los
producing LLSs. The value of f=0.0035 roughly agrees with the observed
values for z$<$4 for PS distribution.  The values for LF distribution
are also plotted in the figure for f=0.0015 and R$_{\rm H}$=500 kpc
which are in reasonable agreement with the observed values.  We use
f=0.0035 for PS distribution and f=0.0015 for LF distribution in what
follows. Here again we note that for z$\sim$ 4 the observed number of
LLS is considerably larger than the model predictions. For f = 0.0035
we estimate that a los with impact parameter $\le$ 100 kpc through
halos with V$_{\rm c} >$ 150 km s$^{-1}$ will have $> 90 \%$
probability of intersecting a cloud. The probability is however less
than 90 $\%$ for f =0.0015. The value of f obtained for PS distribution
is therefore consistent with the lensing data.

The number density evolution of LAASs is displayed in Fig 3 for
R$_{\rm H}$ = 1000 and 500 kpc. As seen, this is sensitive to the
value of halo radius as los producing LAASs does get significant
contribution from the outer parts of the halo. The figure also shows
the observed redshift distribution (taken from Kim et al. 1997)  for
column density $ >10^{13.77} \rm cm^{-2}$, which is also the lower
limit used for the theoretical curves. The distribution is consistent
with the observed distribution for R$_{\rm H}$ =500 kpc, for both PS
and LF distribution. We, however, note that the contribution of the
minihalos is independent of the value of their radius as long as it is
larger than 250 kpc. So we have assumed the minihalo radius to be 250
kpc and the galactic halo radius to be 500 kpc in the rest of the
calculations.

It can be seen from the figures 1, 2 and 3 that the use of LF
distribution without merger is inconsistent with the observations of
the redshift evolution of DLASs, LLSs and LAASs. The observations thus
do indicate a strong evolution in the comoving number density of
galaxies.

The column density distribution function for hydrogen, f(N$_{\rm HI}$),
is defined as the number of absorbing systems per unit column density
per unit redshift path which is defined by X(z) = ${2 \over 3}[ (1+\rm
z)^{3/2} - 1]$ for q$_{o}$ = 0.5. A comparison of the observed column
density distribution with the predictions of our model is presented in
Fig. 4 for the column density range 10$^{12} - 10^{22}$ cm$^{-2}$ at
redshift z = 2.5. The observed data are from Storrie-Lombardi et al,
(2000). A reasonably good match between the theoretical and observed
distributions is obtained for both PS and LF distribution.

\section{\bf {Redshift distribution of heavy element absorption systems }}

For determining the redshift distribution of HEASs, one has to consider
the change in abundance of heavy elements with redshift. It seems
possible that the abundance in galactic halos has been increasing
continuously with time because of in situ star formation (Khare \& Rana
1993; Chiba \& Nath 1997). We therefore assume the abundance of heavy
elements to depend on redshift as Z(z) = Z(0)(1+z)$^{-\delta}$ in the
galactic halos as well as galactic discs and consider several values
for $\delta$. We have also considered another case of an assumed
abundance gradient Z(r, z) = Z(0, z)$\rm e^{-2r/V_{c}}$ for the heavy
elements produced in situ in the galaxies. As there is no star formation 
in the minihalos, the abundance in these halos has been assumed to be 
[Z/H]=-2.5 which has been observed in low column density Ly alpha clouds 
(Songaila \& Cowie, 1996).

The redshift evolution of CIV for N$_{\rm CIV} > 10^{13} \rm cm^{-2}$
is shown in Fig.5 for $\delta$ = 0 and 3 and for $\delta$ = 3, 4 with
an assumed abundance gradient for PS distribution. For LF distribution
lines are shown for $\delta$ = 0 and 3 with an abundance gradient
without merger and  $\delta$ = 3 and 4 with an abundance gradient with
merger. In order to obtain the value of dN/dz from observations, a
maximum likelihood analysis was performed with the data collected from
the literature (references given in Das \& Khare 1999). The data
include components of DLASs also.  These are included in the data,
since most of the components are likely to arise in the galactic
halos. For this analysis we have only used the absorption systems having a
relative velocity w.r.t. the QSOs larger than 5000 km s$^{-1}$. We have
thus excluded any intrinsic absorption systems that may be present in
the QSO spectra. The same holds for any other observational data used
in this paper.  The observed values are plotted in the Fig. 5.
Songaila (1998) has reported observations of CIV toward 13 QSOs. The
value of dN/dz calculated for her data by maximum likelihood analysis
is also plotted in the figure. As seen from the figure, high $\delta$
values (3,4) are indicated by the observations.

We have constructed a sample of observed MgII column densities in
individual components in the absorption systems at redshift z$\sim$0.5
from the literature. We have thus included all the observations of
Lanzetta \& Bowen (1992); Carswell et al. (1991) and Petitjean \&
Bergeron (1990) and performed a maximum likelihood analysis. The
resulting number of systems per unit redshift for lines with N$_{\rm
MgII} > 10^{13.15} \rm cm^{-2}$ seems to require $\delta \ge$ 0.
However, the observed value is for z $\simeq$ 1. More data at higher
redshifts are required to draw any firm conclusion about the value of
$\delta$.

\section{\bf {Column Density Distribution of heavy element lines }}

The density distribution function for MgII, f$_{\rm N_{\rm MgII}}$, is
shown in Fig.6, for the two velocity distribution of the absorbers.
Observed values which are the result of maximum likelihood analysis of
the data sample collected from Lanzetta \& Bowen (1992), Carswell et
al. (1991) and Petitjean \& Bergeron (1990) are also plotted in the
figure.  As evident from the figure, there is a reasonably good
agreement of the model results with the observations with $\delta$=3.
However the column density distribution, being plotted on logarithmic
scale is not very sensitive to the delta values used.

The column density distribution of CIV lines for 2.52$< \rm z <3.78$
has recently been obtained from high resolution Keck observations by
Songaila (1998). This distribution is roughly a power law with a slope
of -1.5 for column densities larger than $6 \times 10^{12} \rm cm^{-2}$
for which her sample is complete. We have plotted model results for the
column density distribution at z=3 for several $\delta$ values in Fig.7
for both PS and LF distribution. Here, as in case of redshift
evolution, results for high $\delta$ values i.e. 3 and 4 are in better
agreement with the observations. However, at lower column densities the
model predicts many more lines than the observed number.  The observed
data, however, may be incomplete below $6 \times 10^{12} \rm cm^{-2}$
as noted by Songaila (1998) and it is possible that the number of small
column density lines may actually be considerably larger. An increase
in the rate of abundance evolution does not serve the purpose, since it
reduces the number of higher column density lines more than that of
lower column density lines.

\section{\bf {Conclusion }}

Pure halo or pure disc will not produce sufficient variety in the
properties of the QSO absorption lines to explain the ensemble of
observed lines. To better understand the structure of the absorbers
and to set constraints on the their properties we have presented a
model which includes minihalo, galactic halos and galactic discs. We
compared various observed distributions of the absorption lines with the
predictions from our model.

The main result of the work presented in this paper is that the
statistical distributions of various classes of absorption systems can
be reasonably well explained by the three component models which are
expected to arise naturally in CDM cosmology. We do not claim the model
to be unique and there could very well be other sets of parameters
which can also explain the observations equally well. We also note that
some of the statistical properties like the redshift distribution of
MgII lines are not very well determined and may change when more
observations become available. However at present we find the
observations to be consistent with the CDM cosmology.

Some of more quantitative results are as follows.

\noindent(1) The ratio of the disc radius to virial radius decreases
with time roughly as (1+z)$^{0.9}$.\newline
\noindent(2) For our assumed baryonic mass of the halo clouds ($ \sim
10^{6} \rm M_\odot $) there are roughly up to a few hundred clouds in
the halos.  Thus roughly a few percent  of the baryonic halo mass is in
the form of a clouds.\newline
\noindent(3) The Lyman alpha lines with column density $>10^{13.77} \rm
cm ^{-2}$ seem to require the radius of the galactic halos to be around
500 kpc assuming this radius to be independent of redshift.\newline
\noindent(4) The column density distribution of HI and MgII is rather
insensitive to variation in various parameters and is in an excellent
agreement with most models, The agreement of the observed column
density distribution of HI with the model results for over 8 orders of
magnitude in column density does strengthen the faith in a unified
model like the one presented in this paper.\newline
\noindent(5) The redshift distribution and column density distribution
of CIV lines does require strong metallicity evolution with redshift as
(1+z)$^{- \delta},\;\delta$ being $\geq$ 3, along with an assumption of
the presence of an abundance gradient in the halo, the abundance
decreasing exponentially with distance. However, we note that CIV lines
mostly arise in clouds embedded in the halos and give us information
regarding the abundance in the halo and not in the discs. The
observations of MgII lines which get contribution from the disc also
are not sufficiently well known to give direct evidence for or against
such an abundance evolution in the discs.\newline
\noindent(6) Observed luminosity function of galaxies is consistent
with the observations of QSO absorption lines only if one assumes that
the comoving number density of galaxies was larger in the past and
reduced with time due to merging.\newline

\section{\bf {Discussion }}

We have presented above a unified model for QSO absorbers where various
classes of QSO absorption systems are produced (apart from the few low
column density Ly $\alpha$ lines produced by the minihalos) by lines of
sight crossing spiral galaxies at different values of impact
parameters. As seen above the model is able to reproduce the observed
redshift and column density distributions of various species of the
absorption systems reasonably well. Several doubts have been expressed
over such a unified scheme in the literature. Bowen et al. (1995)
noticed considerable effect of environment on the absorption lines
produced by galaxies. They, therefore cautioned against constructing
models of generic galactic halos. Diverse morphologies have also been
observed for DLA galaxies (Le Brun et al. 1997; Turnshek et al. 2000;
Kulkarni et al. 2000).

LLS are known to originate in bright galaxies (Steidel et al. 1995).
However, it has been suggested lately that DLASs are produced by low
surface brightness galaxies. High redshift DLASs have not been detected
in Ly $\alpha$ or H $\alpha$ emission except in a few cases, while low
redshift DLASs have been observed to be associated with low surface
brightness galaxies (Turnshek et al. 2000, Pettini et al. 2000).  Rao
and Turnshek (2000) found no evolution in the HI content of DLASs over
the redshift range of 4.0 to 1.0. As a lot of star formation has been
presumably taking place in the bright spiral galaxies over this period
it would have led to a significant  decrease in HI content of these
galaxies. They therefore suggest that DLASs and star forming galaxies
may be different populations. Pettini et al. (1999) showed that DLASs
do not show significant chemical evolution from z$\simeq$3 to 0.3. They also
therefore suggest the DLASs to be different from bright galaxies which
undergo extensive star formation and evolution.  They suggest that the
bright galaxies are under represented in DLA samples precisely for this
reason. The bright galaxies having higher metallicity and therefore
higher dust content will render the QSOs extinct. Savaglio (2000) noted
that the metallicity discussed by Pettini et al. (1999) (being weighted
by column density) is dominated by high column density systems which
may not be uniformly distributed over the entire redshift range.
Restricting to a sample of 50 low column density (N$_{\rm HI} <
10^{20.8} \rm cm ^{-2}$) DLASs which are presumably dust free she found
evidence for chemical evolution. The total sample of 75 QSOs also shows
evolution if dust depletion is corrected for. She found the global
metallicity to change from Z$_\odot$/30 to Z$_\odot$ from z $\sim$ 4.1
to 0.4. It is therefore possible that the observed DLA sample is biased
towards LSB galaxies.

Theoretical studies (Mo, Mao \& White 1998; Mo \& Mao 1999) show that
the halo spin parameter plays an important role in determining the
properties of galaxies forming within halos of dark matter. Halos of
small angular momentum give rise to compact high density systems while
halos of low mass and high angular momentum form discs with a low
surface density of baryons. The former are sites of star formation
while the later dominate the absorption cross section .

Our model does not take into account the presence of dust and assumes
the chemical abundance to be independent of the mass of the galaxies at
a given redshift, though in some of the models we do assume the
chemical abundance to decrease with radial distance, the rate of
decrease being higher for halos with lower circular velocities.
However, even if high metallicity systems are indeed under
represented in the observed DLA sample the rate of chemical evolution
deduced from our model will not change. This is because the rate of
chemical evolution obtained here is from the redshift evolution of the
CIV lines with column density $\ge 10^{13}$ cm$^{-2}$ which is decided
by the los passing only through the halo. Thus the  chemical evolution
deduced in our models is confined to the galactic halos and does not
necessarily reflect a corresponding chemical evolution in the discs.
The thickness of the discs may, however, have to be decreased further
in order to account for the DLASs which are unobserved due to dust
extinction.

\begin{acknowledgements} The authors wish to thank the referee, Joe
Wampler, for his numerous suggestions which improved the manuscript
considerably. This work was partially supported by a grant (No.
SP/S2/013/93) by the Department of Science and Technology, Government
of India.  S.D. and A.S. are supported by C.S.I.R.  fellowships.
\end{acknowledgements}

\noindent {\bf Figure caption:}\\
\noindent {\bf Fig.1:} Redshift distribution of DLASs with column density
greater than 10$^{20.3}$ cm$^{-2}$. Dashed lines marked 1,2,3,4 are for
a=0.0,0.7,0.8,0.9 respectively for PS distribution. Upper and lower
dotted lines are for t = 0.005 R$_d$ and 0.02 R$_d$ respectively for
the PS distribution. All other lines are for t = 0.01 R$_d$. Upper and
lower solid lines are for LF distribution with and without mergers
respectively. Points marked 'x' are observed points from Storrie -
Lombardi et al (1996).\\
\noindent {\bf Fig.2:} Redshift distribution of LLSs. Lower and upper
solid lines are for the PS distribution for R$_{\rm H}$=500 kpc for
f=0.0030 and 0.0035 respectively. Upper and lower dash-dotted lines are
for LF distribution and R$_{\rm H}$=500 kpc with and without mergers
respectively with f=0.0015. Points marked 'x' are observed points from
Storrie - Lombardi et al (1994). \\
\noindent {\bf Fig.3:} Redshift distribution of LAASs with column
density greater than 10$^{13.77}$ cm$^{-2}$. The solid line is the
observed distribution taken from Kim et al (1998). Upper and lower
dotted lines are for R$_{\rm H}$=1000 and 500 kpc respectively with
f=0.0035 for PS distribution. Upper and lower dashed lines  are for
R$_{\rm H}$=500 for LF distribution with and without mergers
respectively with f=0.0015\\
\noindent {\bf Fig.4:} Column density distribution of HI lines at
z=2.5. Points marked 'x' are observed data taken from Storrie-Lombardi et al. (2000) and the dotted line is the best fit to the points. The solid
line is for PS distribution with a=0.9, f=0.0035, R$_{\rm H}$=500 kpc.
Upper and lower dashed lines are for LF distribution with and without
mergers respectively with a=0.9, f=0.0015, R$_{\rm H}$=500 kpc. \\
\noindent {\bf Fig.5:} Redshift distribution for C IV lines with
N$_{\rm CIV}\;>\;10^{13.0}\;\rm cm^{-2}$.  Upper and lower long dashed
lines are for $\delta$=0 and 3 respectively for PS distribution. Upper
and lower short dashed lines are for $\delta$ =3 and 4 with an
abundance gradient respectively for PS distribution. Upper and lower
dotted lines are for $\delta$ = 3 and 4  with an abundance gradient
respectively for LF distribution with mergers.  Upper and lower solid
lines are for $\delta$ = 0 and 3 with an abundance gradient
respectively for LF distribution without mergers. The triangle shows
the observed value for the data set collected from the literature as
described in the text.  and the square represents the value for the
data from Songaila (1998).\\
\noindent {\bf Fig.6:} Column density distribution for MgII lines at z
= 0.5.  Solid line is the observed distribution which is the result of
the maximum likelihood analysis of the data taken from the literature.
Upper and lower dashed lines are for $\delta$= 3 and $\delta$= 3 with
abundance gradient respectively for PS distribution.  Upper and lower
dotted lines  are for $\delta$ = 0 and 3 respectively for LF
distribution without mergers.  Long dashed and dash dotted lines are
for $\delta$ = 0 and 3 respectively for LF distribution with mergers.\\
\noindent {\bf Fig.7:} Column density distribution for C IV lines at z
= 3.0. Long-dashed line is for $\delta$ = 4 and short-dashed line is
for $\delta$=3 with an abundance gradient for PS distribution. Dotted
and long-short dashed line are for $\delta$ = 3 and 4 respectively with
an abundance gradient for LF distribution with mergers. Solid line is
for $\delta$ = 0 for LF distribution without mergers. Short dash-dotted
and long dash dotted lines are for $\delta$ = 3, 4 respectively with an
abundance gradient for LF distribution without mergers. Triangles and
squares are the observed values from Songaila (1998) for z$<$ 3 and z
$\ge$ 3 respectively.\\
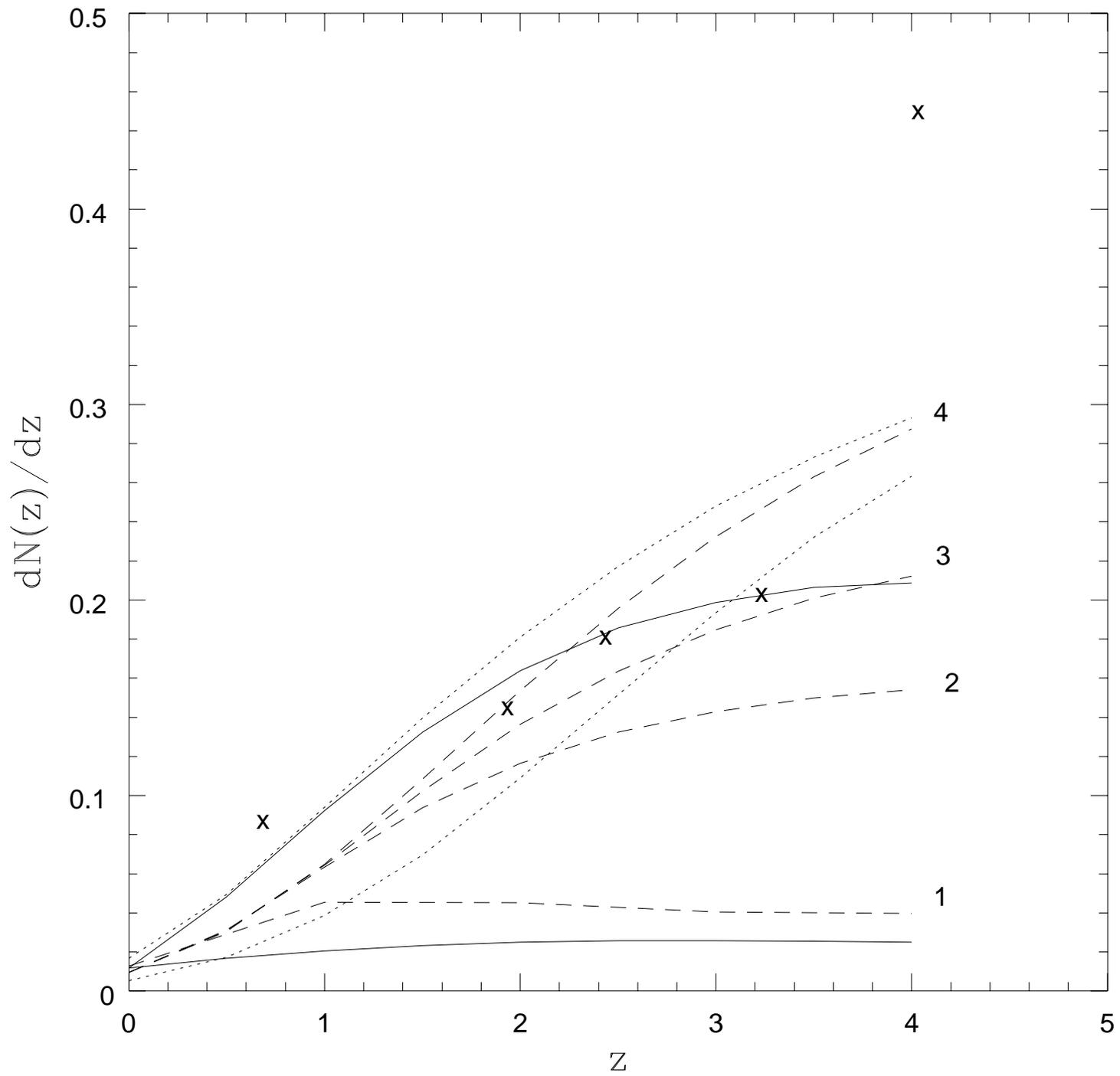
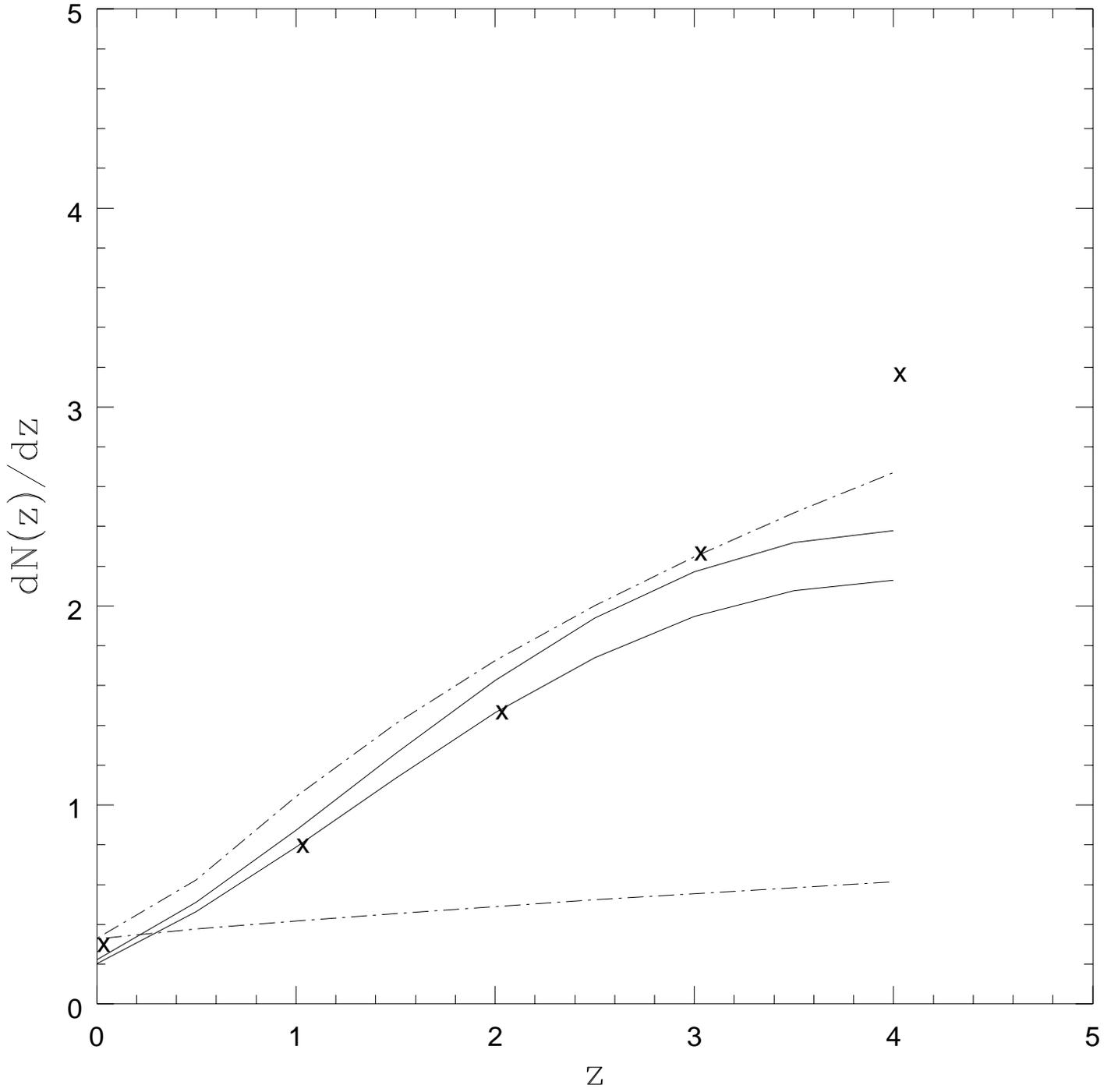
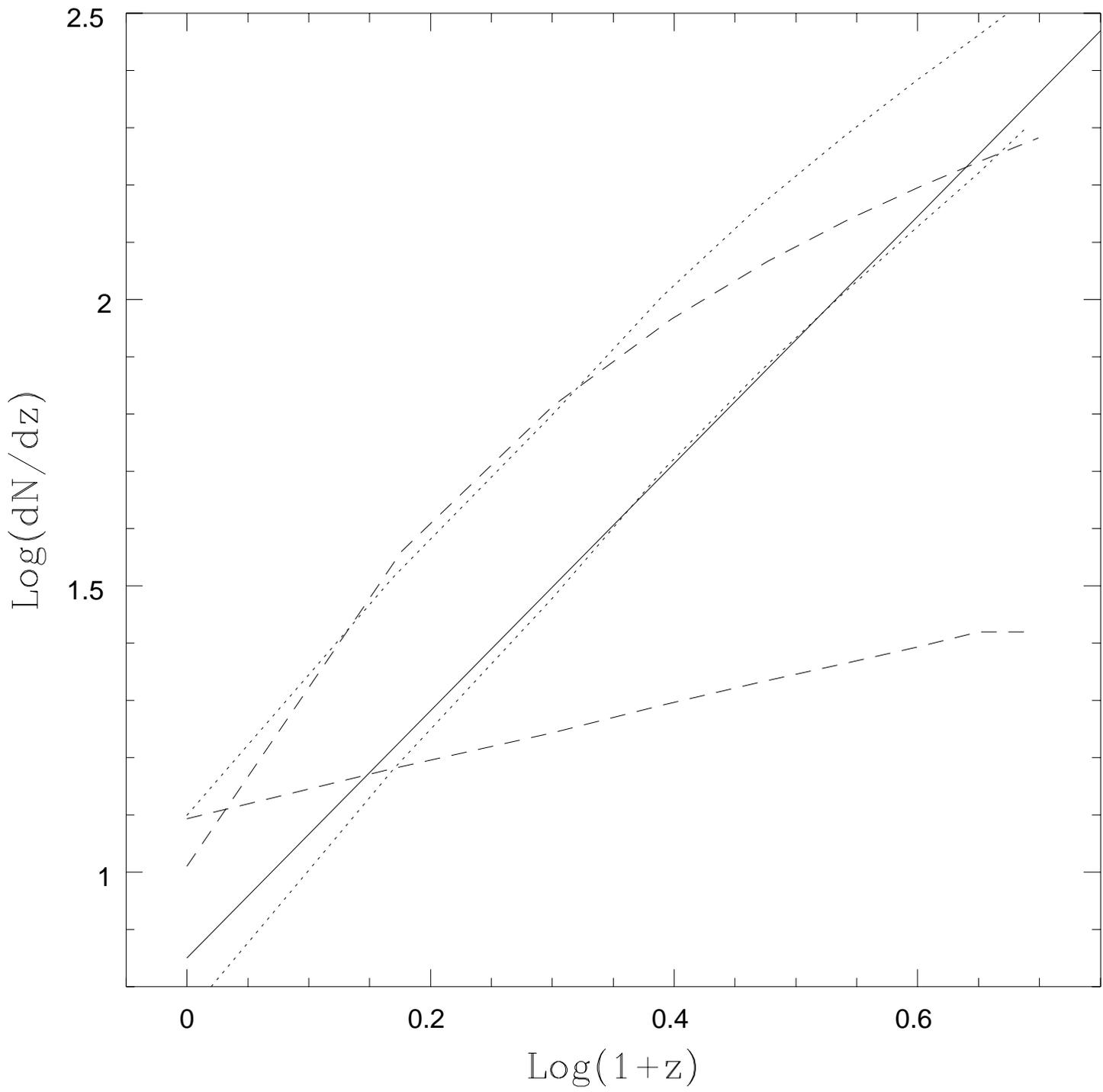
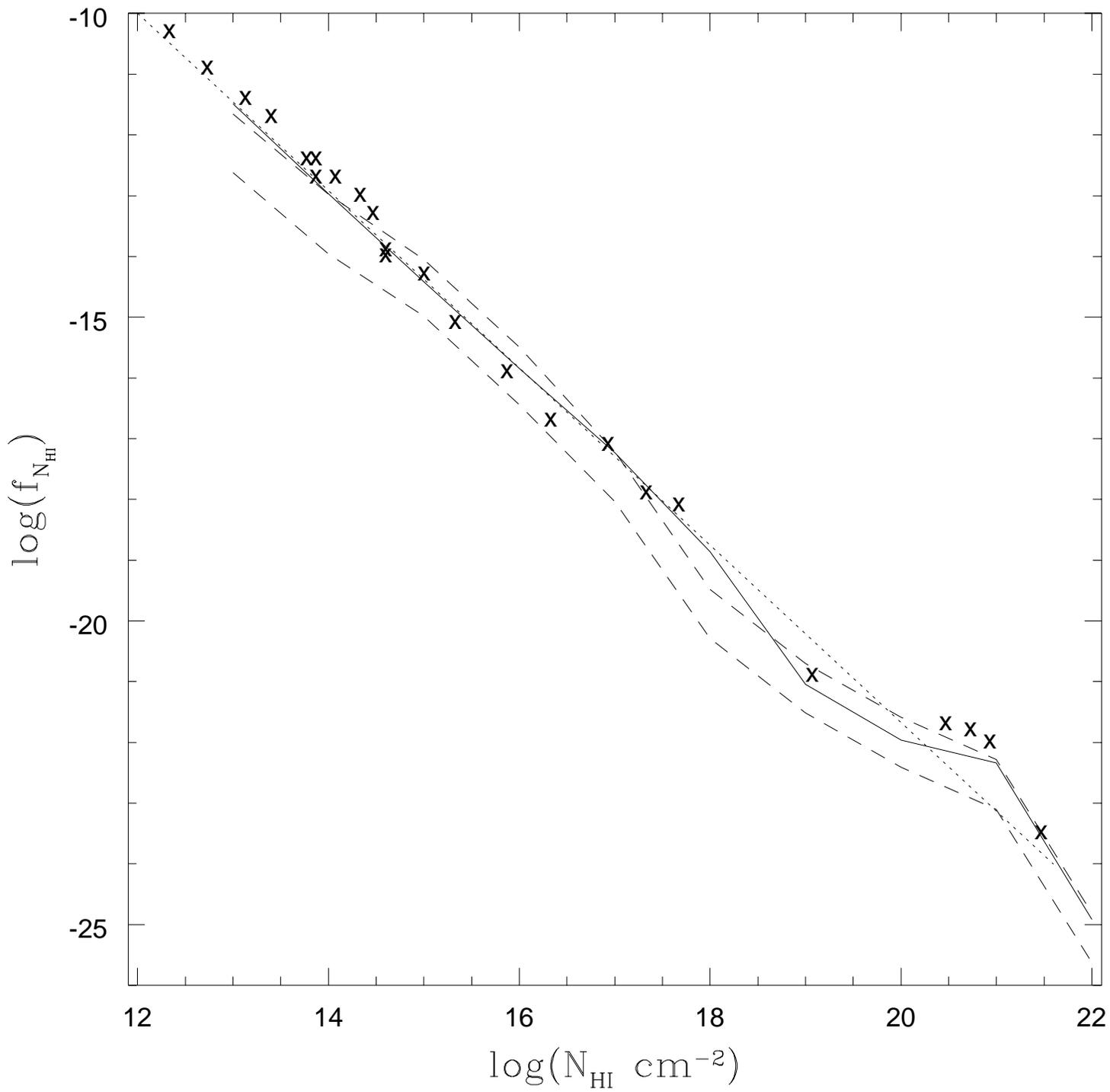
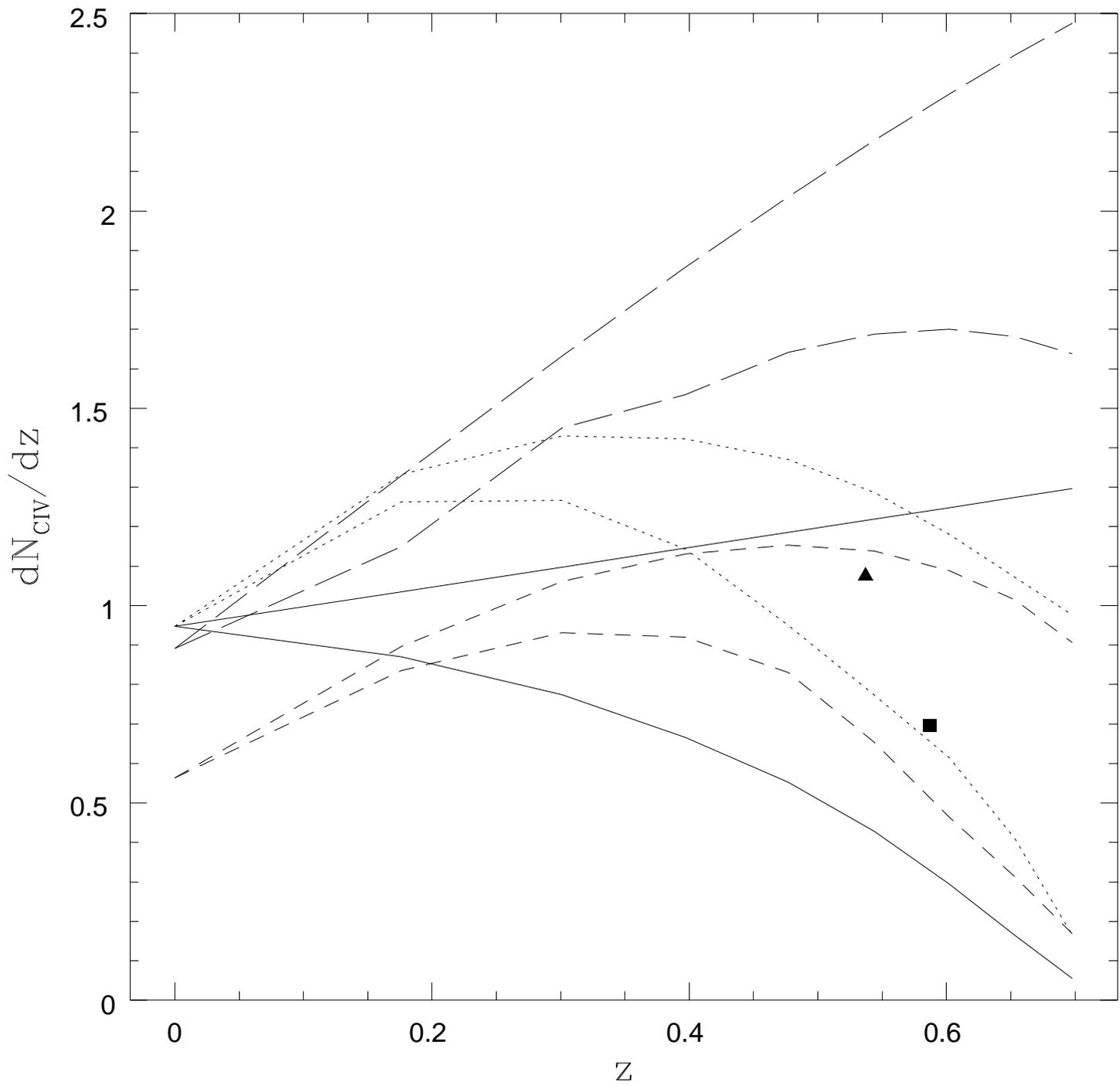
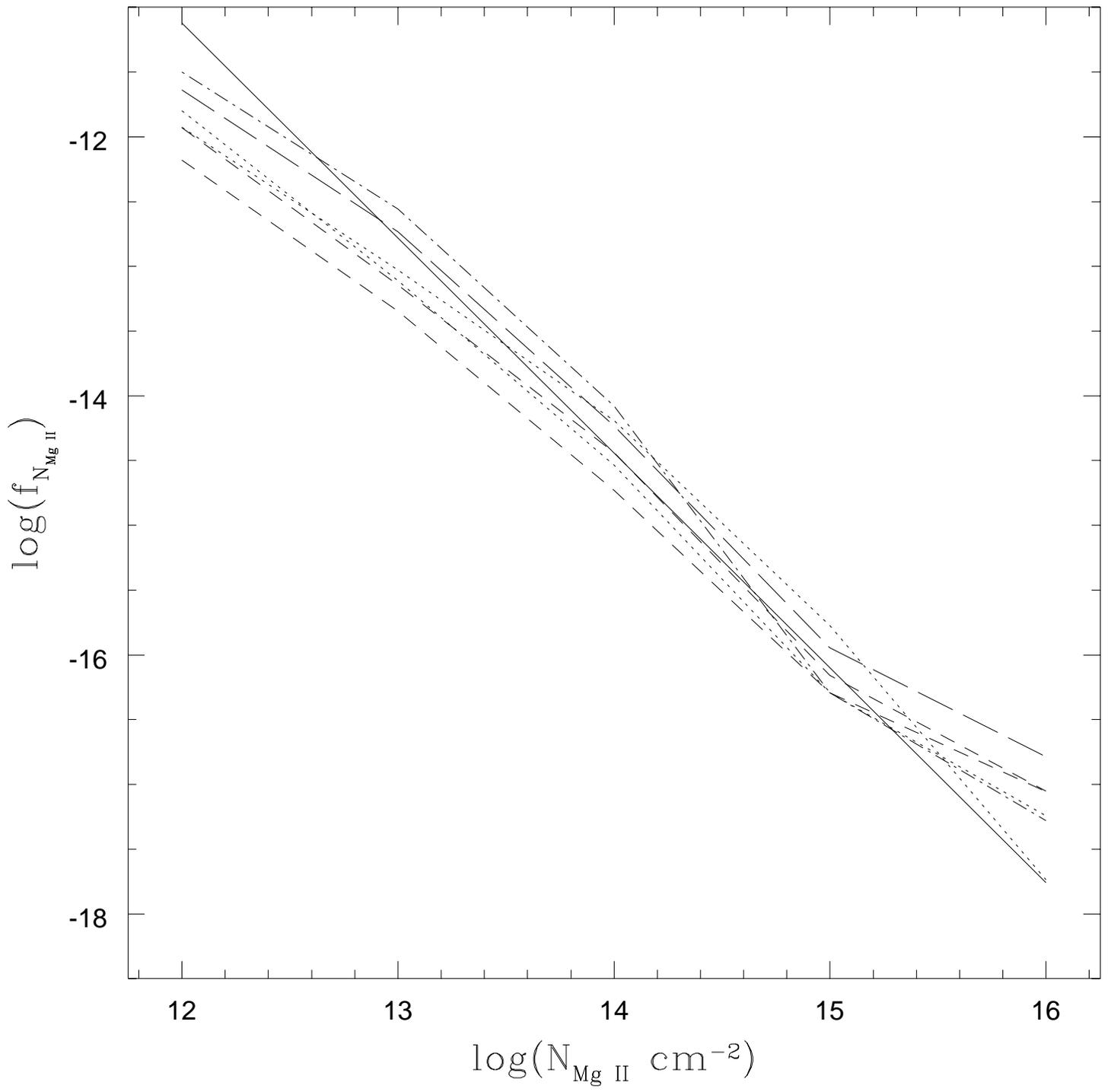
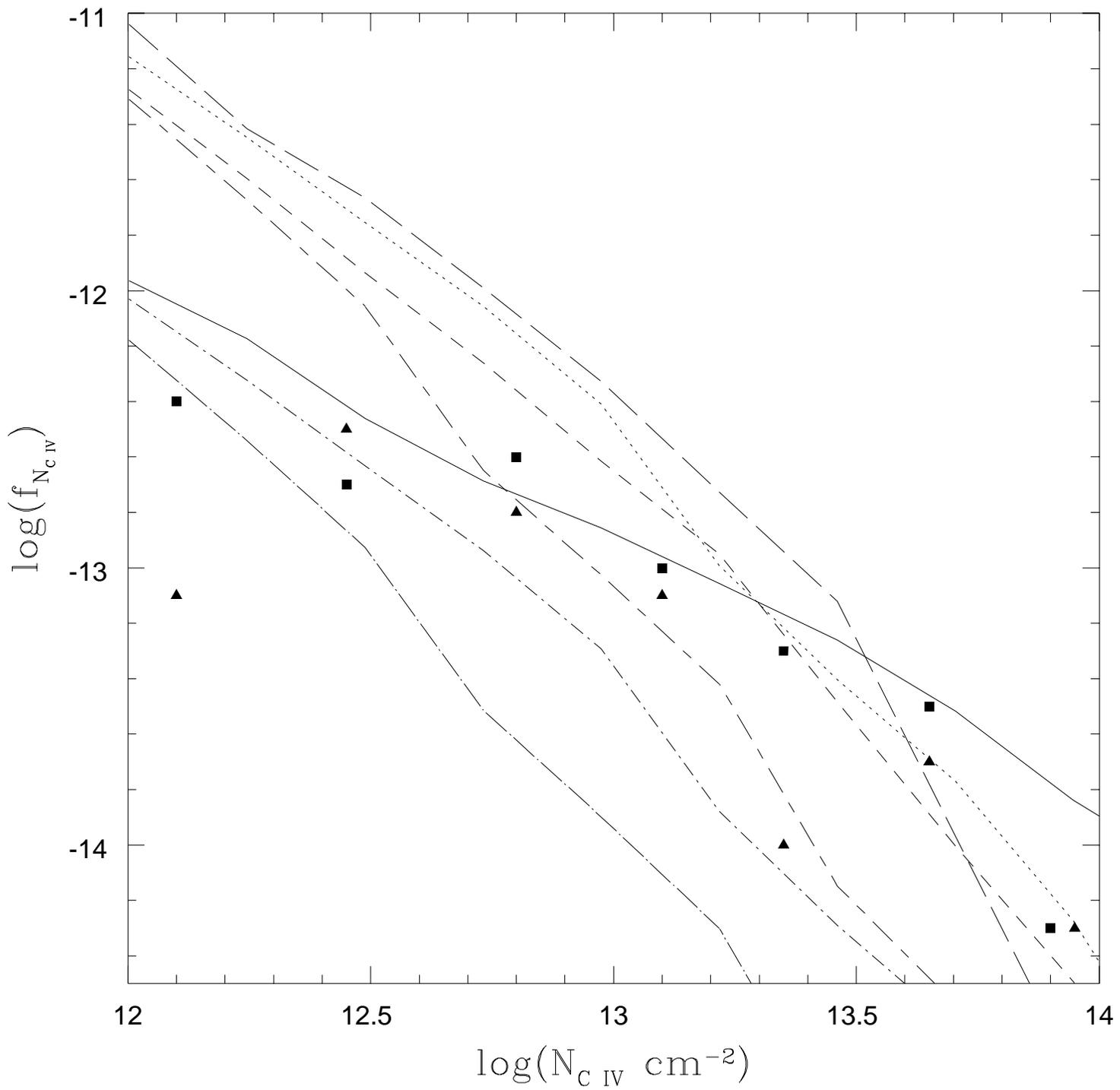
\end{document}